\documentclass[]{article}

\usepackage[english]{babel}
\usepackage{authblk}
\usepackage[dvipsnames,table]{xcolor}
\usepackage{amsmath}
\usepackage{algorithm}
\usepackage[noend]{algpseudocode}
\algnewcommand{\LineComment}[1]{\Statex  \(\triangleright\) #1}
\usepackage{bbm}
\usepackage{tikz}
\usepackage{pgfplots}
\pgfplotsset{compat=1.16}
\usetikzlibrary{patterns.meta,positioning}
\usepackage{tikz-qtree}
\usepackage{multirow}
\usepackage{pifont}
\newcommand{\cmark}{\ding{51}}%
\usepackage{svg}
\usepackage{amsthm}
\usepackage{amsfonts}
\usepackage{url} 
\usepackage{caption}
\usepackage{amssymb}

\newtheorem{theorem}{Theorem}
\newtheorem{lemma}{Lemma}
\newtheorem{definition}{Definition}
\newtheorem{example}{Example}
\newtheorem{remark}{Remark}

\title{BAR Nash Equilibrium and Application to Blockchain Design }
\author[1,2]{Maxime Reynouard\thanks{Corresponding author: \texttt{maximereynouard@gmail.com}}}
\author[2,3]{Rida Laraki}
\author[3]{Olga Gorelkina}
\affil[1]{Nomadic Labs, Paris, France}
\affil[2]{University Dauphine - PSL, Paris, France}
\affil[3]{Mohammed VI Polytechnic University, Rabat, Morocco}
\date{}

\begin{document}
\maketitle 


\begin{abstract}
This paper presents a novel solution concept, called BAR Nash Equilibrium (BARNE) and apply it to analyse the Verifier's dilemma, a fundamental problem in blockchain. Our solution concept adapts the Nash equilibrium (NE) to accommodate interactions among Byzantine, altruistic and rational agents, which became known as the BAR setting in the literature. We prove the existence of BARNE in a large class of games and introduce two natural refinements, global and local stability. Using this equilibrium and its refinement, we analyse the free-rider problem in the context of byzantine consensus. We demonstrate that by incorporating fines and forced errors into a standard quorum-based blockchain protocol, we can effectively reestablish honest behavior as a globally stable BARNE.
\end{abstract}

\section{Introduction}

Security research in the field of Distributed Algorithms (DA) traditionally focuses on fault tolerance. However, the recent proliferation of blockchains has demonstrated that faults are not the only challenge: DA must also resist to self-interested nodes that are neither faulty nor adversarial. In the case of Ethereum, for instance, \cite{unclemaker} has documented instances of miners who violated the prescribed protocol in order to maximize their mining rewards.\footnote{Theoretically, selfish mining attacks were studied in \cite{SelfishMiningMajorityNotEnough,SelfishMiningOptimal,TezosEmmyAttack0,TezosEmmyAttack}; while other types of selfish behaviour were studied in \cite{teutsch2019scalable,DemystifyingIncentives,Arbitrum,Solidus,ShardGameTheory}.} A block creator' maximum extractable value, such as front-running, is another critical dimension of selfish behavior on blockchain. 

While the possibility of self-interested nodes has been acknowledged in the DA literature since at least \cite{BARfaultTolerance}, we are still lacking the theoretical tools to address the fundamental challenge they pose.\footnote{Notable attempts to address robustness to both faulty and selfish nodes include \cite{ByzantineBroadcast} for Byzantine broadcast and \cite{TrapProtocol} for Byzantine consensus.} To propose a universal theoretical framework, our paper turns to game theory, where self-interest---or \emph{rationality}---is a standard assumption. We adapt game theory's central notion of Nash equilibrium \cite{nash1951non} to the scenario  where Byzantine, honest and self-interested (otherwise known as \emph{Byzantine}, \emph{Altruistic}, and \emph{Rational} --- BAR \cite{Robustness_GT-DC, BARfaultTolerance}) nodes coexist.

Our novel solution concept, the BAR-Nash Equilibrium (BARNE), inherits the existence property from \cite{nash1951non}, as we demonstrate in a large class of games including mixed extensions of finite games. Existence is a fundamental property has it allows to always give a prediction of rational agent behavior. BARNE aims at formalizing and generalizing different approaches that were based on intuition rather than methods and that were specific instances of BARNE~\cite{VerifierDilemmaBlog,teutsch2019scalable,Arbitrum,RationalVSByzantines,ByzantineBroadcast}. Two refinements of BARNE, local and global stability, are of  particular relevance for DA design in view of fault tolerance and selfishness tolerance. Stability refers to the robustness of a BARNE to changes in the number of agents of each type. \emph{Locally}  stable BARNE are the strategy profiles that remain BARNE in spite of local perturbations in the numbers of Byzantine and selfish agents. \emph{Globally}  stable BARNE is concerned with the stability of the equilibrium profile for all parameters below certain thresholds in the numbers of Byzantine and selfish agents,in line with the traditional fault tolerance in DA. The notion of BAR-strong equilibrium defined in \cite{Robustness_GT-DC} is an even stricter refinement. We believe both the (stable) BARNE and the BAR-strong equilibrium notions to have their uses, in the vein of the Nash equilibrium and the strong equilibrium of Aumann~\cite{Aumann} in  game theory.

We use BARNE and its refinements to study one of the most pressing blockchain problems, the Verifier's Dilemma, in the context of Quorum Based Consensus Protocols (QBCPs, \cite{tendermint,tenderbake,HotStuff}).\footnote{QBCPs such as Tendermint, Tenderbake, and Hotstuff use adaptations of pBFT \cite{pBFT}, a prominent solution to the Byzantine consensus problem \cite{LamportByzantineGenerals}. QBCPs have the advantage of deterministic block finality, see \cite{TPSvsFinality}.} The Verifier's dilemma arises because multiple agents must verify and validate transactions to maintain blockchain integrity; Since verification is individually costly, it can be rational to forego verifying altogether and  rely on the others' verification effort.\footnote{The Verifier's dilemma is therefore a case of the free-rider problem studied extensively in economics, where it is known to cause a collapse in public good provision, see \cite{hardin1968tragedy,ostrom2008tragedy}. See \cite{VerifierDilemmaBlog} for an excellent informal account of the Verifier's dilemma.}

Our analysis of the Verifier's dilemma shows that following the prescribed strategy of the standard QBCP is almost never a BARNE and is never a stable BARNE.\footnote{When we say that honest behaviour is a BARNE we refer to the behaviour of rational (selfish) agents only.} To restore honest behavior as a stable BARNE, we consider two simple amendments to the classical QBCP. We show that (1) applying monetary penalties for observable deviations from the protocol restores honest verification as a locally stable BARNE and (2) injecting errors \emph{\`a la} \cite{teutsch2019scalable,DemystifyingIncentives} in addition to the penalty results in globally stable BARNE. Obtaining the desired behaviour as a BAR-strong equilibrium requires at least further amendments depending on the discount factor and how communication and transfers operate among the selfish agents. We complete the analysis by studying all other BARNE in the protocol with and without amendments, and show  that free-riding is almost always a globally stable BARNE without the amendments and is never a BARNE in the fully amended protocol. 

A previous work \cite{RationalVSByzantines} formally studied  the free-riding problem in a classical QBCP similar to the one we consider. To solve the Verifier's dilemma, they proposed an amendment where the designer sends personalised, yet correlated recommendations to the agents. Notwithstanding its ingenious design, their amendment is vulnerable: as their construction relies on the exact knowledge of the number of the Byzantine and Selfish agents, the prescribed strategy is not a stable BARNE and is subject to Single Points Of Failure (SPOF). Morevover their analysis only encompassed \textbf{B}yzantine and \textbf{R}ational agents, and not consider the full BAR spectrum.

\paragraph{Our contributions}

First, we extend the fundamental concept of Nash equilibrium and its refinements to the BAR setting:
    \begin{itemize}
        \item We introduce BAR Nash Equilibrium and the notions of local and global stability. We show that \emph{BAR-strong equilibrium $\subset$ globally stable BARNE $\subset$ locally stable BARNE $\subset$ BARNE.} 
        \item We extend Nash-Gliksberg theorem to prove the existence of BARNE under mild conditions.
    \end{itemize}

\noindent Second, we apply BARNE to analyse several proposed solutions to the Verifier's dilemma in QBCPs:
        \begin{itemize}
        \item We show that, in the quorum based consensus protocols that are commonly used at the time of this paper's writing, following the prescribed protocol is almost never a BARNE while the free-riding strategy is a globally stable BARNE.
        \item We propose two realistic amendments of the protocol with which following honestly the prescribed strategy becomes a globally stable BARNE while free-riding is never a BARNE.
        \end{itemize}
    
Our equilibrium notions can apply to study and classify the robustness of other incentive design DA problems as soon as they have the objective to be secure and employed in practice.

\section{BAR Model and Equilibria}\label{sec:BASmodel}
\subsection{Model Primitives}

In line with the distributed algorithms literature, the \emph{agents} are prescribed a protocol $\tau\in T$ (best viewed as a behavioral strategy in an extensive form game), where $T$ denotes an agent's strategy space (all possible deviations of an agent from $\tau$).\footnote{We use \emph{agents} as a catch-all term for processes, participants or nodes of a network.} To measure the ``robustness'' of the protocol, we assume that some agents choose a strategy within $T$ different from $\tau$. We note $N=\{1,2,...,n\}$ of size $n$ the set of all agents, $i\in N$ denotes a single agent and $I \subset N$ a subset. For a (joint) strategy profile $s\in T^n$ of all agents, we note $s_i \in T$  the strategy of agent $i$,  and $s_I \in T^{|I|}$ the sub-profile of agents in $I$. The payoff of agent $i$ when strategy profile $s$ is played is given by $u_i(s_1,...,s_n)$. With a slight abuse of notation, we write $(s_I,s_J,s_i,s_j)=s_{I \cup J \cup \{i,j\}}$, for a strategy profile $s\in T^n$, disjoint subsets $I, J \subset N$ and distinct $i,j \in N \setminus (I \cup J)$. Similarly, in the case of utility functions, we write: $u_i(s)=u_i(s_1,...s_n)=u_i(s_I,s_{N\setminus I})$.

A game is symmetric if $u_i(s_1,...,s_n) =u_{\pi (i)}(s_{\pi (1)},...,s_{\pi (n)})$ for any permutation $\pi$ over $N$. Our definition of BAR Nash equilibrium below and the general existence theorem apply in both symmetric and non-symmetric settings. However, the stability refinements are restricted to symmetric games, owing to our application.

\subsection{The BAR model: three types of agents}

Following \cite{Robustness_GT-DC, BARfaultTolerance}, the \emph{Byzantine--Altruistic--Rational}, or the \emph{BAR} model distinguish three types of agents:

\begin{itemize}
    
    \item Set $F\subseteq N$ of \emph{Byzantine} or \emph{\textbf{F}aulty} agents deviate arbitrarily from $\tau$, including individual and group deviations. Byzantine actions may range from non-strategic and faulty behaviour to collusive and adversarial actions.
    
    \item Set $G\subseteq N$ of \emph{Rational}, \emph{Selfish} or \emph{\textbf{G}ain seeking} agents maximize their payoff in the game. In an incentivised distributed algorithms such as blockchain consensus protocols, gain seeking agents deviate from $\tau$ if it augments their payoff.

    \item Set $H\subseteq N$ of \emph{Altruistic} or \emph{\textbf{H}onest} agents always follow the protocol $\tau$. Honest agents are unable to change the prescribed code or unwilling to do so for the common good's sake.\footnote{Note that actions of Honest and Byzantine types could result from the maximization of \emph{some} utility function. To avoid confusion, we only refer as Rational to those who maximize their direct payoff from the game and as utility to said payoff.} In the rest of the paper, all strategy profiles $s \in T^n$ are assumed to satisfy $s_H=\tau^{|H|}$ in line with this definition.
    
\end{itemize}

Let the cardinal $|F|$ of $F$ be denoted by $f$, $|G|\equiv g,$ and $|H|\equiv h$. Naturally, $\{F, G, H\}$ form a partition of $N$ and thus $f+g+h=n$. 

Classical game theory predominantly focuses on rational agents as its primary actors and employs Nash equilibrium as the prevailing solution concept. On occasion, it delves into local refinements of Nash equilibrium, such as Selten's notion of perfection \cite{Selten1975-SELROT-7} or Myerson's concept of properness \cite{myerson1978refinements}. However, it infrequently employs coalitional refinements like Aumann's strong equilibrium \cite{Aumann}. The next section presents a solution concept which is a version of Aumann's strong equilibrium tailored to the BAR context.

\subsection{BAR-Strong Equilibrium}

\cite{Robustness_GT-DC} and \cite{TrapProtocol} introduced the notion of $(k,t)$-robustness that bears the features of fault tolerance and resistance to coalitional deviations, in the spirit of strong equilibrium \cite{Aumann}. The notion of equilibrium that corresponds to $(k,t)$-robustness is as follows:

\begin{definition} 
A joint strategy profile $s^* \in T^n$ is  a \textbf{$(\bar{f},\bar{g})$ BAR-strong equilibrium}\footnote{Since we assume $n$ is known, there is no need to pin down $h$ once $f$ and $g$ are fixed.} for two given integers $\bar{f}$ and $\bar{g}$ if:
\begin{enumerate}
    \item For all $F\subset N$ such that $f \le \bar{f}$, $s_F \in T^f$ and $i \in N \setminus F$: $u_i(s_F,s^*_{N \setminus F}) \ge u_i(s^*).$
    \item  For all disjoint sets $F,G \subset N$, and strategy profile $s\in T^n$ such that $\ g \le \bar{g}$ and $f\le \bar{f}$, where $s_G \in T^g$ and $s_F \in T^f$, there exists $i \in G$ such that \newline $u_i(s    _F,s_G,s^*_{N \setminus (F \cup G)}) \le u_i(s_F,s^*_{N \setminus F}).$   
\end{enumerate}
\end{definition}

In words, a strategy profile is a \textbf{$(\bar{f},\bar{g})$ BAR-strong equilibrium} if (1) no honest or selfish agent payoff decreases as result of a joint deviation of up to $\bar{f}$ Byzantine agents and (2) no deviation by up to $g$ rational players strictly improves all of the coalition members' payoffs, whatever are the Byzantine's joint strategies. Condition (1) is known as  \emph{$\bar{f}$-immunity} in the DA literature, while condition (2) is equivalent when $g=n$, to a strong Nash equilibrium condition.

Both conditions are fairly restrictive, which prevents BAR-strong equilibrium existence in many games, as pointed out already in \cite{Robustness_GT-DC}. For example, when $\bar{g} \ge 1$, condition (2) implies that $s^*$ is a Nash equilibrium (let $f=0$ and $g=1$). Moreover, when $\bar{g} \ge 2$, condition (2) further implies that no two players can jointly deviate to simultaneously increase their payoff (let $f=0$ and $g=2$). However, in the prisoner's dilemma, these two conditions are incompatible.
Finally, conditions (1) and (2) imply that the equilibrium strategy of the rationals is a best reply to all possible deviations of the Byzantines, a property that seldom exists. 
Strong Nash equilibrium is not popular amongst game theorists because it rarely exists and is not always predictive.\footnote{For example, in the strategic games induced by majoritarian voting methods such as rank voting, plurality or approval voting, a strong equilibrium exists if and only if a Condorcet winner exists and in that case, the Condorcet winner is the unique possible outcome of a strong equilibrium; while in some real instances, the Condorcet winner was not elected with these methods \cite{BalinskiLaraki}} As such, voting theorists developed sophisticated Nash equilibrium refinements to study and compare the outcomes of voting methods \cite{MyersonScoringRules}.

\subsection{BAR-Nash Equilibrium}

Motivated by the above, we now introduce BAR-Nash equilibrium a weaker solution concept that transposes Nash equilibrium  to suit the BAR framework. 

\begin{definition} Given $F$ and $G$, two disjoint subsets of $N$, the joint strategy profile $s^*_G \in T^g$ is
\begin{enumerate}
    \item BARNE at $(F,G)$  if for all  \newline $i\in G$, $s_i^* \in argmax_{s_i\in T}\ min_{s_F \in T^f}\ u_i(s_F,s_i,s_{G\setminus \{i\}}^*,s_H)$. 
    \item BARNE at $(f,g)$ if for all $F$ and $G$ such that $|F|=f$ and $|G|=g$ $s^*_G$ is a BARNE at $(F,G)$.
\end{enumerate}
\end{definition}

\noindent

In contrast to BAR-strong equilibrium outlined in Definition 1, BARNE requires that (1) solely unilateral, and not coalitional deviations are non-profitable; (2) Selfish players best-reply to the worst case scenario as in \cite{Minimax}, and not to \textbf{all} possible faulty deviations; this minimal requirement is very standard in DA; (3) agents know $(F,G)$ or $(f,g)$; this will be relaxed in the next subsection.

Unlike (BAR) strong equilibrium, BARNE is guaranteed to exist under mild conditions, very much like Nash equilibrium.

\begin{theorem}\label{existence} For some given $F$ and $G$, two  disjoint subsets of $N$, noting $H=N \setminus (F \cup G)$, \emph{if} (1) $T$ is a convex compact subset of a topological vector space, (2) any $ i\in G$, $u_i$ is continuous and (3) $t_i \mapsto u_i(s_F,(t_i,s_{G \setminus \{i\}}),s_H)$ is concave for any strategy profile $s \in T^n$, \emph{then} a BARNE exists at $(F,G)$. Moreover, if the game is \textbf{symmetric} then for every $(f,g)$ there exists a \textbf{symmetric BARNE} at $(f,g)$ that is, $\exists \sigma \in T$ s.t. $s_G^*={\sigma}^g$ is a BARNE at $(f,g)$.\end{theorem}

Hence, the existence of a BARNE is guaranteed in particular in mixed extensions of finite games as well as in Euclidean games where the strategy spaces are convex compact and the utility function of each player is jointly continuous and own-strategy concave.

\begin{proof}
Given our assumptions, for $i \in G$ the function: 
\begin{equation*} 
    v_i\colon s_G \in T^g \mapsto \min_{s_F\in T^f} u_i(s_F,s_G,s_H) \in \mathbb{R}
\end{equation*}
is continuous and for any strategy profile $s\in T^n$
\begin{align*}
    w_{i}\colon t_i\in T \mapsto v_i(t_i,s_{G \setminus i}) \in \mathbb{R}
\end{align*}
is concave. Using Theorem 4.7.2 in \cite{RidaMathmeticalFoundationsGT}, we deduce existence of a Nash equilibrium for the game with payoff functions $\{v_i\}_{i\in G}$, which is a BARNE of our game with the payoffs $\{u_i\}_{i\in G}$. 

Moreover, when the game is symmetric, there is a function:
\begin{align*}
    v\colon T \times \Delta_{g-1}(T) & \rightarrow \mathbb{R}\\
    (t,s_{G \setminus i}) & \mapsto v_i(t,s_{G \setminus i})
\end{align*}
where $\Delta_{g-1}(T)$ is the set of degenerate probability distributions on $T$ with support at most $g-1$ and $\pi(s_{G \setminus i})$ is the empirical distribution on $T$ induced by $s_{G \setminus i}$ interpreted as a probability distribution with finite support (it counts how many players used each strategy and normalise). By assumption, $v$ is continuous and concave in the first argument. Hence, if we define the best-reply correspondence:
\begin{align*}
    BR \colon s\in T \mapsto \arg \max_{t \in T} v(t,\pi(s^{g-1})) \in T 
\end{align*}
then $BR$ has non-empty convex-compact values and a close graph and thus has a fixed point.\footnote{By Theorem 4.1 of \cite{RenyExistenceEquilibria}, and Kakutani's theorem if $T$ is locally convex and Hausdorff (see corollary 17.55 in \cite{Aliprantis1994}).} 
\end{proof}

\begin{example}\label{ex:congestion}
The following congestion game provides an example where BARNE exists while the BAR-strong equilibrium does not. Suppose agents are employees connecting to a Virtual Private Network (VPN) for working remotely. Their firm has a slow but fail-proof server $A$ to which they can safely connect and obtain a payoff of $u_A=1$. Connecting to a new, faster server gives a higher payoff of $u_{B1}=2$, but only if $b \le k$ are connected; otherwise the server is of no value, $u_{B2} = 0$. 
In a standard game theoretic setting  with $g=n$, $f=h=0$, there are numerous equilibria where $k$ agents choose $B$ and the rest choose $A$; those could be coordinated by an oracle (company policy). Nevertheless in a BAR setting with $f>0$, if employees $1...k$ connect to $B$, the rest to $A$ and if one of the byzantine was assigned $A$, it can connect to $B$ to crash it, lowering the utility of the users $1...k$. So the protocol is not $1$-immune, and the first condition of robustness in definition 1 fails, so it is not $(f,g)$-strong whenever $f\ge 1$. But worse the second condition prevents any robust equilibrium from existing because of the domination: for a rational, choosing $B$ means taking the risk to be attacked by Byzantine agents crashing $B$; whereas choosing $A$ means taking the chance to miss out on a free spot on $B$ left out by Byzantine agents. However, even in the case where there is only rational and Byzantine agents ($h=0$) BARNE exist: just let $max(k-f,0)$ rationals play $B$ while the others take no risk and play $A$ even if they were assigned $B$ because they would rather be safe than sorry.
\end{example}

\subsection{Locally and Globally Stable BARNE}

Having in mind Blockchain applications, where the number of faulty and selfish players are unknown to the designer, it is natural to wonder whether the (prescribed) strategy being a BARNE at $(f,g)$ implies it being a BARNE at any ($f',g')$ such that $f' \le f$ and $g' \le g$. The answer is no as shown in the next remark. Even though we have the following intermediate result.

\begin{lemma}
For all $g$, $g'$, and $f$ if $\tau^{g}$ is a
BARNE at $(f,g)$, then $\tau^{g'}$ is also a BARNE at $(f,g')$.
\end{lemma}

The proof is immediate if one realizes that when $\tau$ is the equilibrium, to best-reply, it does not matter whether the non-byzantine agents are selfish or honest since in both case they play $\tau$. 

\begin{remark}
However, there exist games where $\sigma \in T \setminus \{\tau\}$ is a BARNE at $(f,g)$ but not at $(f,g')$ with $g'<g$. And there also exist games where $\sigma \in T$ is a BARNE at $(f,g)$ but not at $(f',g)$ with $f'<f$. The game presented in section~\ref{sec:verifier Dilemma} provides examples of both instance, the phenomena is illustrated several times (for instance in figure~\ref{fig:Other_Equ} and figure~\ref{fig:Honest_Equ} respectfully.
\end{remark}

This and the need for Blockchain practitioners for a notion reflecting whether a system can tolerate \emph{up to} a certain amount of byzantines or rationals lead us to the introduction of the next two refinements. The first one is a logical step toward the \emph{up to tolerance} and conveys the possibility that players have an \textit{approximate} knowledge of the $(f,g)$ values. It requires stability of a BARNE with respect to local perturbations around $(f,g)$. 

Motivated by Blockchain applications where all agents are asked to follow the same protocol, our refinements are only defined for symmetric BARNE of the form $\sigma^g$ for some $\sigma \in T$. Consequently, symmetry will be assumed for the rest of the paper.

\begin{definition}
\label{def:loc-stable}
A strategy $\sigma \in T$ constitutes a \textbf{$\delta$-stable BARNE with respect to norm $\Vert \cdot \Vert_\nu$} at $(\dot{f},\dot{g})$, if for all $(f,g)$ such that $\left\Vert (\dot{f},\dot{g})-(f,g) \right\Vert_\nu \le \delta $, $\sigma$ is a symmetric BARNE at $(f,g)$.\footnote{A brief discussion about relevant norms can be found in appendix~\ref{appendix: Norm in simplex}}
\end{definition}

Our second refinement, global stability, is more closely related to the notion of fault tolerance. It conveys the possibility of upper bounds on the numbers of rational and Byzantine agents.

\begin{definition}
A strategy $\sigma \in T$ constitutes a \textbf{globally stable symmetric BARNE} at $(\bar{f},\bar{g})$ if for all disjoint subsets $F$ and $G$ of $N$ such that $f\le \bar{f}$ and $g \le \bar{g}$, $\sigma$ is a BARNE at $(F,G)$.
\end{definition}

If $\sigma$ is a globally stable BARNE at $(\bar{f},\bar{g})$, then it is a $\delta$-stable BARNE at $(\dot{f},\dot{g})$ for all $(\dot{f},\dot{g})$ such that $\dot{f}\leq \bar{f}-\delta$ and $\dot{g}\leq \bar{g}-\delta$. The opposite is not always true. In example \ref{ex:congestion}, no equilibrium would be globally stable, however, when $f>k$, the equilibrium where rational agents all play $A$ is $(f-k)$-stable. This is because even with $f-k$ less Byzantine agents, if one rational chooses $B$ then byzantine can crash it. 
Table \ref{tab:compare_equilibria} illustrates the properties of the various  solution concepts. The more robust is the BAR-strong equilibrium of \cite{Robustness_GT-DC}, but it rarely exists. The less robust is BARNE, but is proven to exist for a large class of games. In between lies globally and locally stable BARNE.

\begin{table*}[ht]
    \rowcolors{2}{white}{gray!25}
    \caption{Properties of different equilibria notions}
    \label{tab:compare_equilibria}
    \noindent
    \makebox[\textwidth]{
    \begin{tabular}{c || c | c | c | c}
       & BAR-strong & BARNE & locally stable BARNE &  globally stable BARNE\\
       \hline
     exist in a large class of games  &  & \cmark &  &  \\
     anti coalition deviations of rationals & \cmark &  &  &  \\ 
      anti individual deviations of rationals & \cmark & \cmark & \cmark & \cmark \\ 
     dominant strategy best-reply wrt Byzantines & \cmark &  &  &  \\
     max-min best-reply wrt Byzantines &  \cmark & \cmark & \cmark &  \cmark  \\
     locally stable &  \cmark &  & \cmark &  \cmark  \\
     globally stable & \cmark & & & \cmark
    \end{tabular}
    }%
\end{table*}

\section{
Analysing some QBCPs using BARNE}\label{sec:verifier Dilemma}

We now use those new concepts to show that free riding is likely to occur in a classical QBCP. Specifically, we show that the prescribed strategy is almost never a BARNE while the free-riding strategy is a globally stable BARNE. We then propose two amendments, allowing the prescribed protocol to become globally stable and the free-riding strategy never a BARNE. 
Appendix~\ref{appendix:BARNE of the initial game} provides a full, step by step, formal analysis of the non-amended stage game (by computing all pure symmetric BARNE for all parameters). Appendix~\ref{appendix:Amended game} provides a shorter analysis for the two versions of the amended game resting on appendix~\ref{appendix:BARNE of the initial game}'s detailed methodology. All of the equilibria results of figures~\ref{fig:Honest_Equ} and~\ref{fig:Other_Equ}) rest on those formal lines of reasoning.

\subsection{A Classical QBCP}

\subsubsection{The endorsement extensive-form game}\label{Sec:QuorumBasedConsensusProtocol}

The Quorum-based consensus protocol studied in our paper (see algorithm \ref{alg:PrescCons}) is similar to that of \cite{RationalVSByzantines}. It describes the protocol aimed at achieving consensus among the agents, i.e., agreement on the new block to add to the chain. The protocol runs in \emph{rounds} and boils down to the following repeated game. Each round, an agent is publicly and randomly\footnote{we suppose a shared and truly random seed} selected to propose a block; the others perform a validity check on the proposal and \emph{endorse} the first valid one they receive. If a \emph{quorum} $Q$ of endorsements is reached then agents consider that consensus is reached, they add the block to the chain, and go to the next level; otherwise a new round starts. It is well known in distributed computing that optimality, meaning the protocol can tolerate the maximum number of Byzantines (up to  $\lceil \frac{n}{2}\rceil - 1$) without effects (invalid blocks being accepted, or valid blocks being rejected), is reached by this protocol for the value $Q=\lfloor \frac{n}{2} \rfloor+1$\footnote{Byzantines can get invalid blocks accepted if $f\ge Q$ and valid blocks rejected if $f>n-Q$, combining these constraints and prioritizing the first gives the results} under the synchronous network assumption\footnote{Network synchrony implies that all messages reach all agents in bounded time and prevents more elaborate attacks.}. For the rest of the paper we will suppose that we have neither $Q \ll n$ (very few endorsements can get a block accepted) nor $n-Q \ll n$ (almost all endorsement are necessary to get a block accepted); those conditions respectfully prevent byzantine to get invalid block accepted or valid block rejected with little effort. 

\begin{algorithm}[ht]
	\caption{Prescribed Consensus Algorithm}
    \label{alg:PrescCons}
	\begin{algorithmic}[1]
        \LineComment{BEGINNING OF A NEW ROUND / PROPOSAL}
		\If{We are the round proposer} 
		    \State \textbf{Create} a new valid block $b$ 
		    \State Propose $b$ on the network
		\EndIf

		\LineComment{ENDORSING}
		\While{\textbf{NOT} (round timeout \textbf{OR} endorsed this round)} 
		    \If{We receive a new block proposal $B$} 
		        \State \textbf{Check} \textbf{validity} of $B$
		        \If{$B$ is \textbf{valid}}
		            \State \textbf{Endorse} $B$
		        \EndIf
		    \EndIf
	    \EndWhile
		    
		\LineComment{DECISION}
		\While{\textbf{NOT} round timeout} 
		    \If{We received $Q$ or more endorsements for $B$}
		        \State add $B$ to our blockchain
		        \State \textbf{GO TO} next level
		    \EndIf
	    \EndWhile
	    
		\State \textbf{GO TO} next round   

	\end{algorithmic} 
\end{algorithm}

This protocol obviously outlines a repeated game, however, to better illustrate our solution concepts but also be closer to what we expect to observe in practice, we focus on the stationary BARNE equilibria (e.g. where the rational agents repeatedly play, iid, the same strategy profile of the stage game). 

Each stage game is a twofold extensive-form game. The proposer plays the proposal game where he decides whether to propose a valid block or not; then the rest of the players play the endorsement game which concerns the rest of the protocol. We focus on the second one since this is the extensive game that concern all players but one, but also that is where most of the difficulty lies. Indeed,  once we consider only that endorsement game, the minimizing strategy for Byzantines consists in always proposing invalid blocks (then endorsing if and only if the block is invalid), whereas honest agents always propose valid blocks then endorse if and only if the block is valid. This means that at least in those cases, we want the valid blocks to be accepted, and the invalid ones rejected, this will suffice to incentivise the rationals to propose valid blocks, hence our focus on the endorsement game.

\subsubsection{Rewards, losses and costs}

Block \emph{validity} typically implies the absence of corrupted data. For crypto-currencies, this includes prohibiting transactions that result in a negative balance for any user. Checking block validity can be arbitrarily costly in the case of smart contracts which can be complex to compute. This creates a free-riding problem where a user can be enticed to deviate from the prescribed protocol by not checking the block validity and endorsing blindly. This saves him the computation cost and he can rely on other users to ensure the rejection of invalid blocks. 

We make the assumption that the rationals take part in the endorsement game simultaneously, i.e., they do not observe each other's endorsements before making their move. If the quorum of $Q$ endorsements is not reached, the block is rejected, and no reward nor loss are incurred; If the quorum was reached, the block is added to the chain, a reward $r_e$ is earned by the endorsers (only those who expressed their endorsement). Designed to compensate the agents who follow the protocol, this reward substantially exceeds the computation costs $c_c$ required to check a block's validity. If the added block was invalid, all agents incur a great loss $L$ that captures the loss of value of their stake due to the failure of blockchain's integrity and reputation.\footnote{Some agents, including the honest types, could refuse to acknowledge the invalid block creating a "fork", but that would lead to its own problems (i.e. risk of insufficient participants to reach the quorum).}
Hence, a rational's payoff is:
\begin{equation}\label{RewardsPlayer}
u =  \mathbbm{1}_{Accepted}\, (\mathbbm{1}_{Endorsed} \, r_e -\mathbbm{1}_{Invalid} \, L) - \mathbbm{1}_{Checked}\, c_c
\end{equation}
where we can assume that $L \gg r_e \gg c_c > 0$.

\begin{figure*}[ht]
\centering
\noindent
\makebox[\textwidth]{
    \begin{tikzpicture}
    \tikzset{style={align=center}}
    \tikzset{grow'=right,level distance=120pt}
    \tikzset{level 1/.style={sibling distance=0mm}}
    \tikzset{level 2/.style={sibling distance=0mm}}
    \tikzset{level 3/.style={sibling distance=0mm}}
    \tikzset{execute at begin node=\strut}
    \tikzset{every tree node/.style={anchor=base west}}
    \Tree [.\node[draw]{Check \\ validity} ;
            \edge node[auto=left]{Y};  [. {Get \\ Validity \\ Information} 
                \edge [dashed] node[auto=left]{Valid}; 
                    [.\node[solid][draw]{Endorse} ; 
                        \edge node[auto=left]{Y}; { $\sigma_h,\ \sigma_{ce}$ }
                        \edge node[auto=right]{N}; { $\sigma_f,\ \sigma_{ce}$ } ]
                \edge [dashed] node[auto=right]{Invalid}; 
                    [.\node[solid][draw]{Endorse} ; 
                        \edge node[auto=left]{Y}; { $\sigma_f,\ \sigma_{c0}$ }
                        \edge node[auto=right]{N}; { $\sigma_h,\ \sigma_{c0}$ } ] ] 
            \edge node[auto=right]{N}; [. {No \\ Information} 
                \edge [dashed] node[auto=left]{}; 
                    [.\node[solid][draw]{Endorse} ; 
                        \edge node[auto=left]{Y}; { $\sigma_e$ }
                        \edge node[auto=right]{N}; { $\sigma_0$ } ] ]  ]
    \end{tikzpicture}
}
\caption{Decision Tree in the Endorsement Game}
\label{tree:decision tree}
\end{figure*}
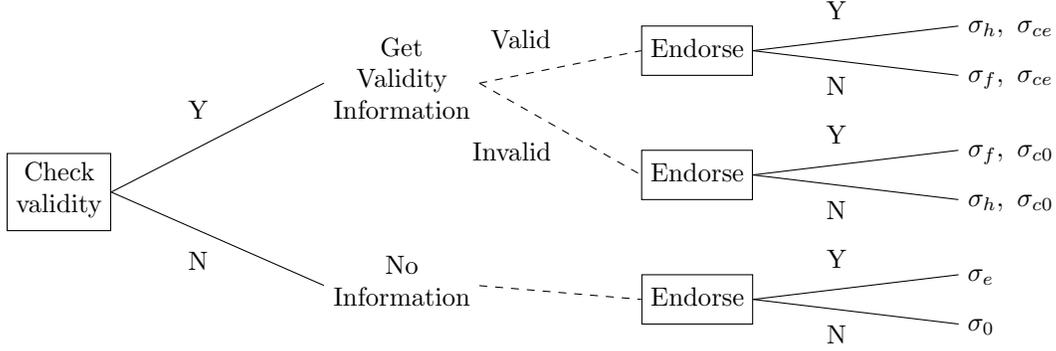

\subsubsection{Strategies in the endorsement game}

The prescribed protocol and the possible deviations an agent could follow form the strategy space $T$. The action space is represented in the tree in figure \ref{tree:decision tree}, boxed nodes and solid edges represent decisions, others are signals. It amounts to six pure strategies represented in the tree's leafs: 
\begin{itemize}
    \item $\sigma_{ce}$: Check validity, endorse unconditionally
    \item $\sigma_{c0}$: Check validity, do not endorse unconditionally
     \item $\sigma_{h}$: Check validity, endorse iff the block is valid. \\
     (The prescribed strategy that Honest or Altruistic agents follow.)
    \item $\sigma_f$: Check validity, endorse iff the block is invalid.\\
    (The minimising strategy of the Byzantine players.)
    \item $\sigma_{e}$: Do not check validity, endorse unconditionally
    \item $\sigma_{0}$: Do not check validity, do not endorse unconditionally
\end{itemize}

Some strategies are \textbf{weakly dominated} for rational players. Failing to endorse a verified valid block only means foregoing the reward $r_e$ when the block is accepted ($\sigma_h$ weakly dominates $\sigma_{c0}$ and $\sigma_f$). Similarly, endorsing an invalid block upon verification only increases the likelihood of the block's acceptance and loss $L$ for the agent ($\sigma_h$ weakly dominates $\sigma_{ce}$ and $\sigma_f$). Thus, most classical Nash refinements (such as Selten's  prefection \cite{Selten1975-SELROT-7} or Myerson's properness \cite{myerson1978refinements}) imply that rationals only choose among the following strategies: endorse without verification $\sigma_e$, no verification and no endorsement $\sigma_0$, or follow the protocol honestly $\sigma_h$ (the prescribed strategy $\tau$ following the notations used in the theory section). As for the Byzantine agents, the most payoff-reducing strategy, $\sigma_f$, in a symmetric BARNE amounts to only endorsing invalid blocks upon verification (and not endorsing valid ones).

\subsubsection{The Byzantine-rational simplex}\label{simplex}

\begin{figure}[ht]
	\centering
	\begin{tikzpicture}
	\begin{axis}[
	xmin = -0.05, xmax = 1.05,
	ymin = -0.05, ymax = 1.05,
	xtick = {0,0.33,0.67,1}, 
	xticklabels ={$0$,$n-Q$,$Q$,$n$},
	ytick = {0,0.33,0.67,1}, 
	yticklabels ={$0$,$n-Q$,$Q$,$n$},
	yticklabel style={rotate=90,anchor=base,yshift=4pt},
	ylabel=rationals $(g)$,
	ylabel shift=-5pt,
	xlabel=byzantines $(f)$,
	legend cell align = {left},
	tick pos=left,
	ytick pos=left,
	]
	
	\addplot[thick, Blue, area legend] coordinates {
		(0,0)
		(1,0)
		(0,1)
	}\closedcycle;
	\addlegendentry{Simplex space}
	
	\addplot[Red, area legend, 
	pattern ={Lines[angle=135,distance = 9pt]},
	pattern color = Red] coordinates {
		(0.67,0)
		(1,0)
		(0.67,0.33)
	}\closedcycle;
	\addlegendentry{Byzantine quorum}
	
	\addplot[Orange, area legend, 
	pattern ={Lines[angle=45,distance = 9pt]},
	pattern color = Orange] coordinates {
		(0.33,0)
		(1,0)
		(0.33,0.67)
	}\closedcycle;
	\addlegendentry{Byzantine veto}
	
	\addplot[OliveGreen, area legend, 
	pattern ={Lines[angle=90,distance = 9pt]},
	pattern color = OliveGreen] coordinates {
		(0,0)
		(0.67,0)
		(0,0.67)
	}\closedcycle;
	\addlegendentry{Honest veto}
	
	\addplot[PineGreen, area legend, 
	pattern ={Lines[angle=0,distance = 6pt]},
	pattern color = PineGreen] coordinates {
		(0,0)
		(0.33,0)
		(0,0.33)
	}\closedcycle;
	\addlegendentry{Honest quorum}
	
	\addplot[thick, Blue, area legend] coordinates {
		(0,0)
		(1,0)
		(0,1)
	}\closedcycle;
	
	\end{axis}
	\end{tikzpicture}
  \caption{The Byzantine-rational simplex, special areas}
  \label{fig:honest_byz_quor_vet}
\end{figure}
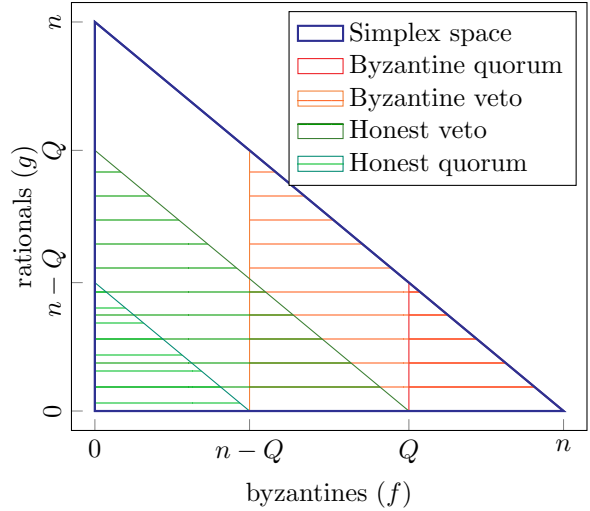

We already mentioned that the parameters $(f,g)$ evolve in a two dimensional simplex scaled up to $n$. Figure \ref{fig:honest_byz_quor_vet} presents this scaled up to $n$ Byzantine-rational simplex. Points $(0,0)$, $(n,0)$ and $(0,n)$ respectively correspond to the cases where all agents are honest, Byzantine, and rational. It is worth noting that four areas of the simplex have special properties: when $f \ge Q$, Byzantines are numerous enough to get any block accepted, we say that we have a Byzantine quorum; similarly, $h \ge Q$ (which is equivalent to $ f+g \le n-Q$) corresponds to an honest quorum. Moreover, when $f > n-Q$ Byzantines are numerous enough to get any block rejected (which is equivalent to $ g+h < Q$: the rest of the players cannot get the quorum by themselves) we have a Byzantine veto; similarly, $h > n-Q$ (which is equivalent to $ f+g < Q$) corresponds to an honest veto.

One can prove that because of the honest veto and quorum, when $(f,g)$ is close to $(0,0)$ (equivalent to $h$ being close to $n$), then $\sigma_e$ strictly dominates the other strategies. Indeed: if we note $p_V$ the proportion of blocks that are valid, since $h$ is close to $n$, then $p_V$ is close to $1$ (honest agents propose valid blocks) so $p_V>0$; moreover since valid blocks are accepted, and invalid ones are rejected, playing $\sigma_e$ yields $u=p_V\ r_e$, whereas playing $\sigma_0$ yields $u=0$ and $\sigma_h$ yields $u=p_V\ r_e - c_c$; this means that $\sigma_e$ yields a higher utility than the other strategy, so it strictly dominates them. 

\subsection{Amending the classical QBCP}

Now we proved that the honest strategy $\sigma_h$ is dominated by the free-riding / blind endorsement strategy $\sigma_e$ for $(f,g)$ close to $(0,0)$ in the classical protocol, we can introduce our two amendments. 

\subsubsection{First amendment}

The most intuitive change is to impose a fine $L_e$ on any agent endorsing an invalid block. Fines are already used in protocols to punish undesirable behaviours. The fines would follow an accusation: an agent broadcast a message with a proof \footnote{The proof broadcasted would be constituted of the invalid block and the signed endorsement of it, additionally one might need to implement a verification game similar to \cite{teutsch2019scalable} to solve the dispute between accuser and accused}. A summary would be included in a block with a debit on the accused frozen stake. Such an implementation does not need the invalid block to be accepted for the fine to be issued. To deter invalid block endorsement, the fine needs to be sufficiently large: $L_e \gg r_e \gg c_c$.

This does not fully resolve the free riding problem as the analysis below will show. For example with $(f,g)=(0,1)$: one rational, the rest are honest. Then since valid blocks are extremely likely (at least $\frac{n-1}{n}$), a rational agent would never play $\sigma_0$ because he wants the endorsement rewards, so valid blocks are accepted, and therefore the rational proposes valid blocks, so all blocks are valid. This means that in this case the rational cannot be fined for playing $\sigma_e$ since there are no invalid blocks to endorse. So $\sigma_e$ still dominating. 

\subsubsection{Second amendment}
It follows that we need to ensure invalid blocks have a minimal probability of being proposed, that is why we propose a second amendment that draws its inspiration from  TrueBit \cite{teutsch2019scalable}. It was proposed, designed and practically implemented for another protocol. Let us suppose that, with a private information, round proposers have a small probability $p_{prop}$ to draw a right to propose an invalid block. Having done so, they wait and see if other agents endorse it. At a later time period, the private information could be revealed to justify the invalid block; its proposer would then receive the usual reward for block creation and the endorsers would be fined for endorsing an invalid block\footnote{The private information could be the cryptographic signature of some data in the last block, which is easily revealed later and authenticated}. As shown in annex, we only require $p_{prop}>\frac{r_e+c_c}{L_e}$ and  since we have $\frac{r_e+c_c}{L_e}\ll 1$, this means that the trap blocks can be quite rare and will not impact the normal functioning of the blockchain.
This condition guarantees that $\sigma_e$ cannot be an equilibrium anymore since it will be dominated by $\sigma_h$ everywhere in the simplex.

\subsection{When is honesty a (stable) BARNE?}\label{sec:Honest_BARNE}

\begin{figure*}[ht]
    \centering
    \noindent
    \makebox[\textwidth]{
	\resizebox{1.6\textwidth}{!}{
\begin{minipage}[t]{0.65\textwidth}
	\caption*{Base protocol, only for $Q \le \frac{n+1}{2}$}%
	\begin{minipage}[b]{\textwidth}
	\centering
		\begin{tikzpicture}
		\begin{axis}[
		xmin = -0.05, xmax = 1.05,
		ymin = -0.05, ymax = 1.05,
		xtick = {0,0.44,1}, 
		xticklabels ={$0$,$Q-1$,$n$},
		ytick = {0,1}, 
		yticklabels ={$0$,$n$},
		yticklabel style={rotate=90,anchor=base,yshift=4pt},
		ylabel=rationals $(g)$,
		ylabel shift=-10pt,
		xlabel=byzantines $(f)$,
		legend cell align = {left},
		tick pos=left,
		ytick pos=left,
		]

		\addplot[Blue] coordinates {
			(0.44,0)
			(0.44,0.56)
		};
		\addlegendentry{Honest BARNE}
		
		\addplot[thick] coordinates {
			(0,0)
			(1,0)
			(0,1)
		}\closedcycle;

		\end{axis}
		\end{tikzpicture}%
	\end{minipage}%
\end{minipage}%
\begin{minipage}[t]{0.65\textwidth}
	\caption*{With 1 amendment}
	\begin{minipage}[b]{\textwidth}
	\centering
		\begin{tikzpicture}
		\begin{axis}[
		xmin = -0.05, xmax = 1.05,
		ymin = -0.05, ymax = 1.05,
		xtick = {0,0.05,0.33,1}, 
		xticklabels ={$0$,$\varepsilon n$,$n-Q$,$n$},
		ytick = {0,1}, 
		yticklabels ={$0$,$n$},
		yticklabel style={rotate=90,anchor=base,yshift=4pt},
		xlabel=byzantines $(f)$,
		legend cell align = {left},
		tick pos=left,
		ytick pos=left,
		]
				
		\addplot[Blue, area legend, 
		pattern ={Lines[angle=45,distance = 9pt]},
		pattern color = Blue] coordinates {
			(0.05,0)
			(0.05,0.95)
			(0.33,0.67)
			(0.33,0)
		}\closedcycle;
		\addlegendentry{Honest locally stable BARNE}
		
		\addplot[thick] coordinates {
			(0,0)
			(1,0)
			(0,1)
		}\closedcycle;
		
		\end{axis}
		\end{tikzpicture}%
	\end{minipage}%
\end{minipage}%
\begin{minipage}[t]{0.65\textwidth}
\caption*{With 2 amendments}
\begin{minipage}[b]{\textwidth}
	\centering
		\begin{tikzpicture}
		\begin{axis}[
		xmin = -0.05, xmax = 1.05,
		ymin = -0.05, ymax = 1.05,
		xtick = {0,0.33,1}, 
		xticklabels ={$0$,$n-Q$,$n$},
		ytick = {0,1}, 
		yticklabels ={$0$,$n$},
		yticklabel style={rotate=90,anchor=base,yshift=4pt},
		xlabel=byzantines $(f)$,
		legend cell align = {left},
		tick pos=left,
		ytick pos=left,
		]
		
		\addplot[Blue, area legend, 
		pattern ={Lines[angle=45,distance = 9pt]},
		pattern color = Blue] coordinates {
			(0,0)
			(0,1)
			(0.33,0.67)
			(0.33,0)
		}\closedcycle;
		\addlegendentry{Honest globally stable BARNE}
		
		\addplot[thick] coordinates {
			(0,0)
			(1,0)
			(0,1)
		}\closedcycle;
		
		\end{axis}
		\end{tikzpicture}%
	\end{minipage}%
\end{minipage}%
	}
 }
	\caption{Areas of the Byzantine-rational simplex where 
		the honest strategy $\sigma_h$ is a BARNE}
	\label{fig:Honest_Equ}
\end{figure*}

In Figure \ref{fig:Honest_Equ}, we can see the areas of the simplex where the honest strategy is a BARNE (the precise proof is presented in the appendix). As mentioned in the figure (left graphic), for the baseline protocol the equilibria are not stable (any change in $f$ would break it) and the strategy can only be a BARNE if $Q\le \frac{n+1}{2}$. This is because for the BARNE to hold, we need the rationals to be pivotal in the acceptance of invalid blocks (this forces $f=Q-1$) and we need the valid blocks to be able to be accepted (so we need to be outside the byzantine veto: $f \le n-Q$). Those constraints put together force $Q\le \frac{n+1}{2}$. This is a problem, since even under the favorable condition of a synchronous network we would like $Q>\frac{n}{2}$, leaving only the case where $n$ is even and $Q= \frac{n+1}{2}$; in the more realistic asynchronous setting with $Q=\lfloor \frac{2n}{3} \rfloor +1$ this cannot work. Here is a list of the issues with ``following the prescribed strategy'' for the classical BFT protocol (without amendments): 
\begin{itemize}
    \item A precise value for $f$: this BARNE is not stable.
    \item Unrealistic: currently in real world blockchains, most users are honest ($f$ is small). 
    \item Since we also need that $f \le n-Q$, this forces $Q \le \frac{n+1}{2}$ which is an unacceptable limit on $Q$.\footnote{most byzantine consensus protocol set $Q$ around $\frac{2n}{3}$ because of more advanced attacks in the asynchronous setting}
    \item The number of Byzantine agents is critical: if one more user turns Byzantine, the blockchain fails. This is a Single Point Of Failure (SPOF).
\end{itemize}
The last point sheds a light on the conflict between the rules of distributed computing design and game theory. In game theory, a rational user will not pay a cost for the common good if he does not have agency on the outcome of the game. But at the same time, distributed computing strives to limit any single agent's agency to avoid SPOFs. 

As we can see in figure \ref{fig:Honest_Equ} (middle graphic), with one amendment, the honest BARNE becomes locally stable, we still need to be outside the byzantine veto ($f \le n-Q$) but can now tolerate a big set of values for $f$. However, when $f$ nears $0$, we loose the equilibrium. This is because the fines need to be a credible threat so that the loss they create is greater than the checking cost $c_c$ that could be saved with the blind endorsement $\sigma_e$, so we need $ f \ge \varepsilon \ n$ with $\varepsilon = \frac{c_c}{L_e} \ll 1$.

With the two amendments, the honest BARNE becomes globally stable, only the constraint of being outside the byzantine veto needs to be respected $f \le n-Q$, otherwise no rewards can be earned, so agents might as well save the checking cost and do nothing ($\sigma_0$). This is optimal because consensus cannot function properly in the rest of the simplex: as explained, the Byzantine veto ($f>n-Q$) is a theoretical bound that cannot be further optimised.

\subsection{Unstability of a previous solution}

A recent paper \cite{RationalVSByzantines} studied a quorum-based game similar to ours but only in the particular case where $f \le min(Q,n-Q)$ and $h=0$ (no agent is honest). They designed a clever mechanism to solve the verifier's dilemma (using our language, following their protocol is a BARNE). To do so, they use a correlation device which fairly assigns roles to agents. Their mechanism has the flavour and the failures of the honest equilibrium in the original protocol because it makes sure that $Q-1$ agents (the $f$ Byzantines, and $Q-1-f$ among the rationals) will be recommended to play $\sigma_e$ (the free-riding strategy). The rest of the players ($Q$ of them) are recommended to use $\sigma_h$ (the honest strategy). Consequently, this BARNE is an SPOF (an additional byzantine would break the protocol); it relies on the precise knowledge of $f$ so it is an unstable BARNE.

\subsection{On the other BARNEs of the games}

\begin{figure*}[ht]
	\centering
    \noindent
    \makebox[\textwidth]{
\resizebox{1.6\textwidth}{!}{
	\begin{minipage}[t]{0.65\textwidth}
		\caption*{$\sigma_{0}$ BARNE, base protocol}%
		\begin{minipage}[b]{\textwidth}
			\centering
			\begin{tikzpicture}
			\begin{axis}[
			xmin = -0.05, xmax = 1.05,
			ymin = -0.05, ymax = 1.05,
			xtick = {0,0.33,0.67,1}, 
			xticklabels ={$0$,$n-Q+2$,$Q$,$n$},
			ytick = {0,0.33,1}, 
			yticklabels ={$0$,$n-Q+2$,$n$},
			yticklabel style={rotate=90,anchor=base,yshift=4pt},
			ylabel=rationals $(g)$,
			ylabel shift=-5pt,
			legend cell align = {left},
			tick pos=left,
			ytick pos=left,
			]

			\addplot[Red, area legend, 
			pattern ={Lines[angle=45,distance = 9pt]},
			pattern color = Red] coordinates {
				(0,0.33)
				(0.33,0)
				(0.67,0)
				(0.67,0.33)
				(0,1)
			}\closedcycle;
			\addlegendentry{Cold start BARNE}
			
			\addplot[thick] coordinates {
				(0,0)
				(1,0)
				(0,1)
			}\closedcycle;

			\end{axis}
			\end{tikzpicture}%
		\end{minipage}%
	\end{minipage}%
	\begin{minipage}[t]{0.65\textwidth}
		\caption*{$\sigma_{0}$ BARNE, with 1 amendment}%
		\begin{minipage}[b]{\textwidth}
			\centering
			\begin{tikzpicture}
			\begin{axis}[
			xmin = -0.05, xmax = 1.05,
			ymin = -0.05, ymax = 1.05,
			xtick = {0,0.33,1}, 
			xticklabels ={$0$,$n-Q+2$,$n$},
			ytick = {0,0.33,1}, 
			yticklabels ={$0$,$n-Q+2$,$n$},
			yticklabel style={rotate=90,anchor=base,yshift=4pt},
			legend cell align = {left},
			tick pos=left,
			ytick pos=left,
			]
			
			\addplot[Red, area legend, 
			pattern ={Lines[angle=45,distance = 9pt]},
			pattern color = Red] coordinates {
				(0,0.33)
				(0.33,0)
				(1,0)
				(0,1)
			}\closedcycle;
			\addlegendentry{Cold start BARNE}
			
			\addplot[thick] coordinates {
				(0,0)
				(1,0)
				(0,1)
			}\closedcycle;
			
			\end{axis}
			\end{tikzpicture}%
		\end{minipage}%
	\end{minipage}%
	\begin{minipage}[t]{0.65\textwidth}
		\caption*{$\sigma_{0}$ BARNE, with 2 amendments}%
		\begin{minipage}[b]{\textwidth}
			\centering
			\begin{tikzpicture}
			\begin{axis}[
			xmin = -0.05, xmax = 1.05,
			ymin = -0.05, ymax = 1.05,
			xtick = {0,0.33,1}, 
			xticklabels ={$0$,$n-Q+2$,$n$},
			ytick = {0,0.33,1}, 
			yticklabels ={$0$,$n-Q+2$,$n$},
			yticklabel style={rotate=90,anchor=base,yshift=4pt},
			legend cell align = {left},
			tick pos=left,
			ytick pos=left,
			]
			
			\addplot[Red, area legend, 
			pattern ={Lines[angle=45,distance = 9pt]},
			pattern color = Red] coordinates {
				(0,0.33)
				(0.33,0)
				(1,0)
				(0,1)
			}\closedcycle;
			\addlegendentry{Cold start BARNE}
			
			\addplot[thick] coordinates {
				(0,0)
				(1,0)
				(0,1)
			}\closedcycle;
			
			\end{axis}
			\end{tikzpicture}%
		\end{minipage}%
	\end{minipage}%
}
}
    \noindent
    \makebox[\textwidth]{
\resizebox{1.6\textwidth}{!}{
	\begin{minipage}[t]{0.65\textwidth}
		\caption*{$\sigma_{0}$ BARNE, base protocol}%
		\begin{minipage}[b]{\textwidth}
			\centering
			\begin{tikzpicture}
			\begin{axis}[
			xmin = -0.05, xmax = 1.05,
			ymin = -0.05, ymax = 1.05,
			xtick = {0,0.05,0.67,1}, 
			xticklabels ={$0$,$\varepsilon n$,$Q$,$n$},
			ytick = {0,0.67,1}, 
			yticklabels ={$0$,$Q$,$n$},
			yticklabel style={rotate=90,anchor=base,yshift=4pt},
			ylabel=rationals $(g)$,
			ylabel shift=-5pt,
			xlabel=byzantines $(f)$,
			legend cell align = {left},
			tick pos=left,
			ytick pos=left,
			]

			\addplot[Red, thick] coordinates {
				(0.05,0.62)
				(0.67,0)
			};
			\addlegendentry{Not a BARNE}
			
			\addplot[Orange, area legend, 
			pattern ={Lines[angle=45,distance = 9pt]},
			pattern color = Orange] coordinates {
				(0,0)
				(0.67,0)
				(0,0.67)
			}\closedcycle;
			\addlegendentry{Honest Veto BARNE}
			
			\addplot[Blue, area legend, 
			pattern ={Lines[angle=90,distance = 6pt]},
			pattern color = Blue] coordinates {
				(0,0.67)
				(0.67,0)
				(1,0)
				(0,1)
			}\closedcycle;
			\addlegendentry{Breakdown BARNE}
			
			\addplot[OliveGreen, area legend, 
			pattern ={Lines[angle=135,distance = 9pt]},
			pattern color = OliveGreen] coordinates {
				(0,0)
				(0.05,0)
				(0.05,0.95)
				(0,1)
			}\closedcycle;
			\addlegendentry{Threatless BARNE}

			\addplot[Red,thick] coordinates {
				(0.05,0.62)
				(0.67,0)
			};
			
			\addplot[thick] coordinates {
				(0,0)
				(1,0)
				(0,1)
			}\closedcycle;

			\end{axis}
			\end{tikzpicture}%
		\end{minipage}%
	\end{minipage}%
	\begin{minipage}[t]{0.65\textwidth}
		\caption*{$\sigma_{e}$ BARNE, with 1 amendment}%
		\begin{minipage}[b]{\textwidth}
			\centering
			\begin{tikzpicture}
			\begin{axis}[
			xmin = -0.05, xmax = 1.05,
			ymin = -0.05, ymax = 1.05,
			xtick = {0,0.05,0.1,1}, 
			xticklabels ={$0$,$\varepsilon_1 n$,$\varepsilon_2 n$,$n$},
			xticklabel style={rotate=55,anchor=north east,yshift=0pt},
			ytick = {0,0.67,1}, 
			yticklabels ={$0$,$Q$,$n$},
			yticklabel style={rotate=90,anchor=base,yshift=4pt},
			xlabel=byzantines $(f)$,
			xlabel shift=-9pt,
			legend cell align = {left},
			tick pos=left,
			ytick pos=left,
			]
			
			\addplot[OliveGreen, area legend, 
			pattern ={Lines[angle=45,distance = 6pt]},
			pattern color = OliveGreen] coordinates {
				(0,0)
				(0.05,0)
				(0.05,0.62)
				(0.1,0.57)
				(0.1,0.9)
				(0,1)
			}\closedcycle;
			\addlegendentry{Threatless Equilibrium}
			
			\addplot[thick] coordinates {
				(0,0)
				(1,0)
				(0,1)
			}\closedcycle;

			\end{axis}
			\end{tikzpicture}%
		\end{minipage}%
	\end{minipage}%
	\begin{minipage}[t]{0.65\textwidth}
	\caption*{$\sigma_{e}$ BARNE, with 2 amendments}%
	\begin{minipage}[b]{\textwidth}
		\centering
		\begin{tikzpicture}
		\begin{axis}[
		xmin = -0.05, xmax = 1.05,
		ymin = -0.05, ymax = 1.05,
		xtick = {0,1}, 
		xticklabels ={$0$,$n$},
		ytick = {0,1}, 
		yticklabels ={$0$,$n$},
		yticklabel style={rotate=90,anchor=base,yshift=4pt},
		xlabel=byzantines $(f)$,
		legend cell align = {left},
		tick pos=left,
		ytick pos=left,
		]
		
		\addplot[thick, area legend] coordinates {
			(0,0)
			(1,0)
			(0,1)
		}\closedcycle;
		\addlegendentry{Not a BARNE}
		
		\end{axis}
		\end{tikzpicture}%
	\end{minipage}%
\end{minipage}%
}
}
	\caption{Areas of the Byzantine-rational simplex where 
		$\sigma_0$, or $\sigma_e$ are BARNEs}
	\label{fig:Other_Equ}
\end{figure*}

In figure \ref{fig:Other_Equ}, one can see that the blind endorsement strategy $\sigma_e$ where the rational agents do not check the block and endorse them is a BARNE almost everywhere in the baseline protocol, and that the fines alone did not suffice to prevent it to be a BARNE when $f$ is very small. However, with both amendments, since  the prescribed strategy $\sigma_h$ dominates $\sigma_e$ over the entire simplex, the free riding strategy $\sigma_e$ cannot be a BARNE anymore.

The cold start equilibrium with the "do nothing" strategy $\sigma_0$ (no block checking, no endorsement) remains quite present over the simplex, it is inadvisable to try and suppress it through repressive measures as agents playing $\sigma_0$ are indistinguishable from honest agents having a temporary fault (network issue or other...). Solutions taking us out of our model could be considered such as proposing empty blocks when one notice that blocks are not accepted anymore (this would nullify the checking cost $c_c=0$, changing our model and making $\sigma_h$ weakly dominate $\sigma_0$ which would make the cold start equilibrium unlikely).

\subsection{Conditions for a BAR-strong equilibrium} 

Looking at the twice-amended game, we need to clarify how agents would collude before we can discuss a possible BAR-strong equilibrium. Here we could assume that rational agents can engage in private communication and can transfer utility by redistributing their gains. This means that only a single member needs to increase its utility without affecting the others for the coalition to beat the honest equilibrium. In our case, a coalition can have one of their member check the block, and communicate its validity to the others before endorsing it. This can lead to centralization (the gains from the coalition are super-linear, attracting more and more members) which in turns could lead to vulnerability (a byzantine in  the coalition can get it to endorse an invalid block). In our case one might wonder (1) whether the honest strategy is BAR-strong equilibrium in the static endorsement game? (2) if yes, does this hold in the repeated game?

To answer (1) we can refer to the results of \cite{TrapProtocol} which established a necessary condition for having a BAR-strong equilibrium: the game needs a rewarded baiting strategy which consists in a way to get a reward for betraying the coalition. Here the fine can serve as a way to betray the coalition. But to meet the reward part of the condition we need to twist our first amendment to pay (part of) the fine to the accuser. In doing so the condition becomes sufficient. Indeed, the coalition shatters as it is in the participants interest to betray it to earn the fine from the other participants. 

About (2): even though a BARNE of a static game is a BARNE of any associated discounted repeated games, it is not true for a BAR-strong equilibrium. Indeed, depending on the discount rate, the momentary gain from betraying the coalition for the fine could not suffice to compensate the recurring gain from the coalition. Further work is needed to see whether we can establish new conditions for a BAR-strong equilibrium to hold in the repeated game.

\section{Conclusion} 

We have shown that the standard QBCP is vulnerable to the Verifier's dilemma: following the protocol is not a locally stable BARNE, and is almost never a BARNE, while free-riding is a  globally stable BARNE. Introducing fines and trap blocks can prevent free riding from being a BARNE and turn honest behavior into a globally stable. Moreover, contrary to former solutions, our protocol's prescription forms a locally and even a globally stable BARNE. 
 
On the practicality of the two amendments: fines, are not novel for blockchain protocols especially QBCP. For example, Tenderbake~\cite{tenderbake}, the protocol for Tezos, fines agents who propose two blocks at the same level; meanwhile forced errors, the analog of our trap blocks, have been used in practice. For example \cite{teutsch2019scalable,DemystifyingIncentives} informed the design of real-world solutions within Layer 2 with a similar purpose of providing incentives for data verification. Therefore, we argue that achieving a globally stable BARNE is practically implementable if one adopts the two amendments.

At the methodology level, our article introduces the concept of BARNE and proves its existence in a large class of games that go well beyond the scope of the QBCP application presented here. Combined with the local and global stability refinements as well as the BAR-strong equilibrium of \cite{AbrahamOptimal}, those game theoretical tools can help to design more robust mechanisms in the BAR setting.

\section*{Acknowledgments}
We would like to thank Lacramioara Astefanoaei and Eugen Zalinescu for their helpful comments and feedback.

\bibliographystyle{unsrt}
\bibliography{biblio}


\appendix

\clearpage

\section{BARNE of the initial game} \label{appendix:BARNE of the initial game}

Here we will go through the methodology we used to compute the symmetric pure BARNE everywhere in the simplex. We already explained the game formed by the consensus, we will now establish a table similar to the normal form of a game: a payoff table depending on the agent's strategy and the unknown state of the block. Then we will aggregate those payoffs into an expected payoff according to a belief distribution over the block state and derive inequalities comparing the strategies. Finally we will explain how, why and where in the simplex the different strategies can be BARNE

\subsection{Payoffs}

We adopt the following notation. In tables, the expected payoff from all three strategies will be placed in each cells in this manner:
\begin{center}
\begin{tabular}{||c|c||}
    \hline \hline
    \color{Red}{$u(\sigma_0)$} & \color{Green}{\underline{$u(\sigma_e)$}} \\
    \hline
     \multicolumn{2}{||c||}{\color{Orange}{\boldmath{$u(\sigma_h)$}}} \\
    \hline \hline
\end{tabular}
\end{center}

Positions correspond to strategies: top-left for $\sigma_0$, top-right for $\sigma_e$ and the bottom for $\sigma_h$. The color and font express how payoffs from different strategies are ordered: {\color{Green}\underline{$u$}}$>${\color{Orange}\boldmath{$u$}}$>${\color{Red}$u$}. A rational's payoff from different strategies, depending on the block state space are all reported in table~\ref{tab:resultPivotal}. Block-state space has two dimensions: $Validity \times Acceptance$. Validity is determined by the proposer's action and unknown by all except him; we use the notation $Validity = \{V,\ I\}$ for \emph{Valid, Invalid}. Acceptance also is unknown during the endorsement phase: it depends on the decision phase and we use $Acceptance = \{A,\ R,\ P\}$ for \emph{Accepted} (the block is accepted independently of the agent's action); \emph{Rejected} (the block is rejected independently of the agent's action): \emph{Pivotal} (the agent determines acceptance or rejection, e.g., the block receives exactly $Q-1$ other endorsements).

\begin{table}[ht]
\caption{Payoff depending on block acceptance (rows), block validity (columns), and chosen strategy (sub-cells).}
\label{tab:resultPivotal}
\centering
\begin{tabular}{||c||c|c||c|c||}
    \hline
     & \multicolumn{2}{|c||}{Valid} & \multicolumn{2}{c||}{Invalid} \\
    \hline\hline
    \multirow{2}{*}{Accepted} & {\color{Red}$0$} & {\color{Green}\underline{$r_e$}} & {\color{Orange}\boldmath{$-L$}} & {\color{Green}\underline{$r_e-L$}} \\
    \cline{2-5}
                              & \multicolumn{2}{c||}{{\color{Orange}\boldmath{$r_e-c_c$}}} & \multicolumn{2}{c||}{{\color{Red}$-L-c_c$}} \\
    \hline\hline
    \multirow{2}{*}{Rejected} & {\color{Green}\underline{$0$}} & {\color{Green}\underline{$0$}} & {\color{Green}\underline{$0$}} & {\color{Green}\underline{$0$}} \\
    \cline{2-5}
                              & \multicolumn{2}{c||}{{\color{Red}$-c_c$}} & \multicolumn{2}{c||}{{\color{Red}$-c_c$}} \\
    \hline\hline
    \multirow{2}{*}{Pivotal} & {\color{Red}$0$} & {\color{Green}\underline{$r_e$}} & {\color{Green}\underline{$0$}} & {\color{Red}$r_e-L$} \\
    \cline{2-5}
                              & \multicolumn{2}{c||}{{\color{Orange}\boldmath{$r_e-c_c$}}} & \multicolumn{2}{c||}{{\color{Orange}\boldmath{$-c_c$}}} \\
    \hline\hline

\end{tabular}
\end{table}

\subsection{Computing the expected payoff 
}

Since we look for BARNE we suppose that agents know $f$, $g$ and $h$. From this and the equilibria they might be in, we suppose they would share a belief distribution over the block state, given in table~\ref{tab:probas}, we can use this to compute the expected payoffs.
Here,  $p_V$, $p_I$, $p_A$, $p_R$, $p_P$ denotes the probabilistic belief that the block is respectively \emph{valid} or \emph{invalid} and \emph{accepted}, \emph{rejected}, 
 or \emph{pivotal} (depending on other agents' endorsements). Finally we  denote $p_{i,j},\ i\in\{A,\ R,\ P\},\ j \in \{V,\ I\}$ the probability beliefs that the block is in an intersection. Following properties hold: $p_i= \sum_{j \in A}p_{ji}, \forall i \in V$; $p_j= \sum_{i \in V}p_{ji}, \forall j \in A$; $\sum_{i \in V, j\in A}p_{ij}=1$; but $p_{ij}=p_i\ p_j$ is not necessarily true since block state are not necessarily independent over the two spaces (i.e., in case of honest veto and honest quorum with $f>0$, among the $p_ij$, only $p_{AV}$ and $p_{RI}$ are greater than $0$ which would be incompatible with the independence)

\begin{table}
\centering
\caption{ Probability-belief distributions in block state space}
\label{tab:probas}
\begin{tabular}{c|c c}
 & $p_V$ & $p_I$ \\
 \hline
$p_A$ & $p_{AV}$ & $p_{AI}$ \\
$p_R$ & $p_{RV}$ & $p_{RI}$ \\
$p_P$ & $p_{PV}$ & $p_{PI}$
\end{tabular}
\end{table}

Using those beliefs, we can compute the expected payoffs from the three strategies and aggregate them for comparison in a table looking like this: 
\begin{center}
\begin{tabular}{c|c }
    $\mathbbm{E}\left(u(\sigma_0)\right)$ & $\mathbbm{E}\left(u(\sigma_e)\right)$  \\
    \hline
     \multicolumn{2}{c}{$\mathbbm{E}\left(u(\sigma_h)\right)$ }
\end{tabular}
\end{center}

Once computed we reach the expected payoffs from table~\ref{tab:Expected}

\begin{table}
\caption{ Expected payoffs from the three strategies}
\centering
\label{tab:Expected}
\begin{tabular}{ p{4.2cm} | p{4.2cm} }
    \centering $ - p_{AI} \, L$ & $  (p_A+p_P) \, r_e - (p_{AI}+p_{PI}) \, L $  \\
    \hline
     \multicolumn{2}{c}{$ (p_{AV}+p_{PV}) \, r_e  - c_c - p_{AI} \, L$ }
\end{tabular}
\end{table}

Three sets of inequalities allow us to analyse the best responses for rational  agents (the symbol $\lesseqqgtr$ reads "less, equal, or greater" and allows us to proceed with calculations with equivalence of the unresolved inequalities): 
\begin{equation}\label{ineq:participation_1}
\begin{split}
    \mathbbm{E}\left(u(\sigma_h)\right) & \lesseqqgtr \mathbbm{E}\left(u(\sigma_0)\right) \\
   (p_{AV}+p_{PV}) \, r_e  - c_c - p_{AI} \, L & \lesseqqgtr - p_{AI} \, L \phantom{+(p_{AV}+p_{PV}) \, r_e  - c_c }\\
    (p_{AV}+p_{PV}) \, r_e & \lesseqqgtr c_c
\end{split}
\end{equation}

\begin{equation}\label{ineq:noLazy_1}
\begin{split}
    \mathbbm{E}\left(u(\sigma_h)\right) & \lesseqqgtr \mathbbm{E}\left(u(\sigma_e)\right) \\
    \makebox[0pt][r]{\text{$(p_{AV}+p_{PV}) \, r_e  - c_c - p_{AI} \, L $}}& \lesseqqgtr \makebox[0pt][l]{\text{$(p_A+p_P) \, r_e - (p_{AI}+p_{PI}) \, L$}}\\
    \makebox[0pt][r]{\text{$p_{PI} \, L  $}}& \lesseqqgtr \makebox[0pt][l]{\text{$(p_{AI} + p_{PI}) \, r_e + c_c$}}
\end{split}
\end{equation}

\begin{equation}\label{ineq:0_vs_e}
\begin{split}
    \mathbbm{E}\left(u(\sigma_0)\right) & \lesseqqgtr \mathbbm{E}\left(u(\sigma_e)\right) \\
    \makebox[0pt][r]{\text{$- p_{AI} \, L$}} & \lesseqqgtr \makebox[0pt][l]{\text{$(p_A+p_P) \, r_e - (p_{AI}+p_{PI}) \, L$}}\\
    \makebox[0pt][r]{\text{$p_{PI} \, L$}}  & \lesseqqgtr \makebox[0pt][l]{\text{$(p_A+p_P) \, r_e$}}
\end{split}
\end{equation}

\subsection{Computing the symmetric pure BARNE of the voting game}

In order to look for all symmetric pure BARNE, we proceed strategy by strategy by looking at how much endorsements each block would get depending on their validity: Valid blocks always get the $h$ endorsements from honest agents, and they get $h+g$ if rationals play $\sigma_e$ or $\sigma_h$. Meanwhile invalid blocks always get the $f$ endorsements from byzantine agents, and they get $f+g$ if rationals play $\sigma_e$. From this we can deduce proprieties on beliefs $p_{i,j}$:

\begin{itemize}

    \item  First $\sigma_e$. Note that in both inequalities~\ref{ineq:noLazy_1} and ~\ref{ineq:0_vs_e} the only thing that could prevent $\sigma_e$ from yielding the best payoff is if $p_{PI}>0$. So to break the equilibrium we need the $\sigma_e$ playing rationals to be pivotal for invalid blocks, so without deviation invalid blocks need to receive exactly $Q$ endorsements. At the same time, since rationals play $\sigma_e$, invalid blocks get $f+g$ endorsements. This outlines three cases depending on the comparison of $f+g$ and $Q$:
    
    \begin{itemize}
    
        \item Case of $f+g<Q$: then $p_{PI}=0$ and $\sigma_e$ is a BARNE. And $f+g<Q$ (which is equivalent to $ h > n-Q$), so we only need to be in the honest veto setting for the strategy to be a BARNE. Hence the name of the equilibrium.
        
        \item Case of $f+g>Q$: then $p_{PI}=0$ and $\sigma_e$ is a BARNE. This time, invalid blocks are accepted with at least $Q+1$ endorsements. So no single rational can change the block acceptance and the chain is failing, hence the name Breakdown equilibrium.
        
        \item Case of $f+g=Q$: this where difficulty lies. This time rational agents will be pivotal for invalid blocks, so we will have $p_{PI}=p_I$, this also means $p_{AI}=0$, so $p_{A}=p_{AV}$. For $\sigma_e$ to be a BARNE, inequality \ref{ineq:0_vs_e} therefore say that we need $(p_{AV} + p_{P})r_e \ge p_I (L-r_e)$. So, either $f=0$ to ensure $p_{I}=0$ since no byzantine is here to propose invalid blocks (the right part is null), or $h+g \ge Q $ (which is equivalent to $f \le n-Q$) so that $p_{PV}>0$ or $p_{AV}>0$, this leave a chance to the left part to be greater than the right. Note that the former condition encompasses the latter (since if $f=0$, then $h+g=n>Q$), so we need $h+g \ge n$ (equivalent to $f \le n-Q$). This gives $p_{V}=p_{AV}+p_{PV}$, and since we have $p_{I}=p_{PI}$ this gives $p_{A}+p_{P}=1$. So inequality \ref{ineq:0_vs_e} becomes $p_{I} \le \frac{r_e}{r_e+L}$.
        
        Additionally, using $p_{PI}=p_{I}$ (so $p_{AI}=0$) which we established, inequality \ref{ineq:noLazy_1}, say that we need $p_{I} \le \frac{c_c}{L-r_e}$ which is more restrictive than the previous inequality $p_{I} \le \frac{r_e}{r_e+L}$ since $L \gg r_e \gg c_c$. So now we only need: $f\le n-Q$ and $p_{I} \le \frac{c_c}{L-r_e}$. Since all blocks would be accepted, rationals would propose valid blocks to get the rewards without the loss, so only byzantine would propose invalid blocks: $p_I=\frac{f}{n}$. So now we only need: $f\le n-Q$ and $f \le \frac{c_c}{L-r_e} n$ (which is $\ll n$), for more readability we consider the second restriction to encompass the first one (we only suppose that $Q$ is not too close to $n$ which is sensible) and therefore we keep only this necessary condition $f \le \frac{c_c}{L-r_e} n$. But if we want to be perfectly rigorous, we could use the condition $f\le \varepsilon  = min(n-Q, \frac{c_c}{L-r_e} n)$. This means the Byzantines are too few to constitute a significant threat for the rational agents. That's how we obtain this threat-less equilibria.
    \end{itemize} 
    
    \item Second $\sigma_0$. In this case, let us show that $f < Q$ by contradiction. The inverse ($f \ge Q$) would mean that invalid blocks are accepted ($p_{PI}=0$) and that at least some are proposed ($p_{A} \ge p_{AI}>0$); with inequality \ref{ineq:0_vs_e} we would get $\mathbbm{E}\left(u(\sigma_0)\right)  < \mathbbm{E}\left(u(\sigma_e)\right)$ which would break the supposed equilibrium. So $ f < Q$. Similarly, we cannot have $h \ge Q-1$ since this means valid blocks are accepted or pivotal ($p_{AV}+p_{PV}=p_{V}$) and some are proposed in non-negligible proportions as we supposed that we do not have $Q \ll n$; since $c_c \ll r_e$ this would mean $p_{AV}+p_{PV} > \frac{r_e}{c_c}$ and inequality \ref{ineq:participation_1} does not hold for $\sigma_0$ to be an equilibrium. So now $f < Q$ and $h< Q-1$, which means all blocks are rejected. This means $p_{AV}+p_{PV}=0$ and $p_{A}+p_{P}=0$ guaranteeing that inequalities \ref{ineq:participation_1} and \ref{ineq:0_vs_e} are satisfied for the equilibrium to hold. So under those conditions, $\sigma_0$ is a BARNE called the cold start.

    \item Finally we consider the honest strategy $\sigma_h$. When rational agents are playing honestly, valid blocks are getting $h+g$ endorsements, and invalid ones are getting $f$. From inequality~\ref{ineq:noLazy_1}, we can see that we need $p_{PI}>0$ for $\sigma_h$ to have a chance against $\sigma_e$ because $c_c>0$. This means that we need $Q-1$ endorsements on invalid blocks, forcing $f=Q-1$, this is sufficient to guarantee the inequality is in favor of $\sigma_h$ as $L \gg r_e \gg c_c$  and we supposed that we do not have $Q\ll n$ so $p_PI$ is non-negligible. At the same time, from inequality~\ref{ineq:participation_1}, we can see that wee need $p_{AV}+p_{PV}>0$ for $\sigma_h$ to have a chance against $\sigma_0$. This means that we need at least $Q$ endorsements on valid blocks, forcing $h+g \ge Q$ (which is equivalent to $ f \le n-Q$), this is sufficient as once again $Q$ is not negligible. Combined with $f=Q-1$, this gives us $2Q-1 \le n$ so $Q \le \frac{n+1}{2}$. So if $Q\le \frac{n+1}{2}$ and $f=Q-1$ then $\sigma_h$ is a BARNE called the honest BARNE. 
\end{itemize}

\section{Amended game}\label{appendix:Amended game}
\subsection{Only with the fines over invalid endorsements}

The fines change agents' payoffs as reported in Table \ref{tab:resultPivotalAmend}.

\begin{table}
\caption{Rational agents' payoffs depending on block acceptance (rows), block validity (columns), \& agent's own strategy (sub-cells)}
\label{tab:resultPivotalAmend}
\centering
\begin{tabular}{||c||c|c||c|c||}
    \hline
     & \multicolumn{2}{|c||}{Valid} & \multicolumn{2}{c||}{Invalid} \\
    \hline\hline
    \multirow{2}{*}{Accepted} & {\color{Red}$0$} & {\color{Green}\underline{$r_e$}} & {\color{Green}\underline{$-L$}} & {\color{Red}$r_e-L-L_e$} \\
    \cline{2-5}
                              & \multicolumn{2}{c||}{{\color{Orange}\boldmath{$r_e-c_c$}}} & \multicolumn{2}{c||}{{\color{Orange}\boldmath{$-L-c_c$}} }\\
    \hline\hline
    \multirow{2}{*}{Rejected} & {\color{Green}\underline{$0$}} & {\color{Green}\underline{$0$}} & {\color{Green}\underline{$0$}} & {\color{Red}$-L_e$} \\
    \cline{2-5}
                              & \multicolumn{2}{c||}{{\color{Red}$-c_c$}} & \multicolumn{2}{c||}{{\color{Orange}\boldmath{$-c_c$}}} \\
    \hline\hline
    \multirow{2}{*}{Pivotal} & {\color{Red}$0$} & {\color{Green}\underline{$r_e$}} & {\color{Green}\underline{$0$}} & {\color{Red}$r_e-L-L_e$} \\
    \cline{2-5}
                              & \multicolumn{2}{c||}{{\color{Orange}\boldmath{$r_e-c_c$}}} & \multicolumn{2}{c||}{{\color{Orange}\boldmath{$-c_c$}}} \\
    \hline\hline

\end{tabular}
\end{table}

The expected payoffs of the three strategies become:

\begin{center}
\begin{tabular}{ p{1.2cm} | p{5.5cm} }
    $ - p_{AI} \, L$ & $  (p_A+p_P) \, r_e - (p_{AI}+p_{PI}) \, L - p_I \, L_e$  \\
    \hline
     \multicolumn{2}{c}{$ (p_{AV}+p_{PV}) \, r_e  - c_c - p_{AI} \, L $ }
\end{tabular}
\end{center}

The payoff inequalities $u(\sigma_h) \lesseqqgtr u(\sigma_0)$, $u(\sigma_h) \lesseqqgtr u(\sigma_e)$, $u(\sigma_0)  \lesseqqgtr u(\sigma_e)$ become, respectively:

\begin{equation}\label{ineq:participation_2}
    (p_{AV}+p_{PV}) \, r_e  \lesseqqgtr c_c
\end{equation}

\begin{equation}\label{ineq:noLazy_2}
    p_{PI} \, L  + p_I \, L_e  \lesseqqgtr (p_{AI} + p_{PI}) \, r_e + c_c 
\end{equation}

\begin{equation}\label{ineq:0_vs_e_3}
    p_{PI} \, L + p_I\, L_e  \lesseqqgtr (p_A+p_P) \, r_e
\end{equation}

From there we can lead the same reasoning as before to find the BARNE. 

\begin{itemize}
    \item First for $\sigma_e$, invalid blocks get $f+g$ endorsements, valid ones get $h+g$ and given the fines, we will need invalid blocks to be rare for the equilibrium to hold (for example, inequality \ref{ineq:0_vs_e_3} forces $p_I \le \frac{r_e}{L_e}$, so $f$ needs to be small $f \ll n$. this means that $h+g \approx n$, so valid blocks are accepted (we supposed we do not have $n-Q \ll1$). Since valid blocks are accepted, rationals propose valid blocks to get rewards without suffering $L$ or $L_e$, so  $p_{V}=p_{AV} \approx 1$. We then distinguish three cases:

    \begin{itemize}
        \item When $f+g<Q$ then invalid blocks are rejected so $p_{I}=p_{RI}$ and $p_{PI}=p_{AI}=0$. Inequality \ref{ineq:noLazy_2} yields $p_{I}\le \frac{c_c}{L_e}$, and inequality \ref{ineq:0_vs_e_3} yields $p_{I} \le p_{A} \frac{r_e}{L_e} \approx\frac{r_e}{L_e}$ since $p_{A}= p_{V} \approx 1$. The first condition is much stronger since $c_c \ll r_e \ll L_e$, so we only need $f \le \frac{c_c}{L_e} n$ for the equilibrium to hold
        \item When $f+g>Q$, then invalid blocks are accepted so $p_{I}=p_{AI}$, $p_{PI}=p_{RI}=0$ and so $p_A=1$. Inequality \ref{ineq:noLazy_2} yields $p_{I}\le \frac{c_c+p_I\ r_e}{L_e}$ which unfolds as $p_{I}\le \frac{c_c}{L_e-r_e}$, and inequality \ref{ineq:0_vs_e_3} yields $p_{I} \le p_{A} \frac{r_e}{L_e} = \frac{r_e}{L_e}$. The first condition is much stronger since $c_c \ll r_e \ll L_e$ so we only need $f \le \frac{c_c}{L_e-r_e} n$ for the equilibrium to hold.
        \item Finally when $f+g=Q$, then invalid blocks are pivotal so $p_{I}=p_{PI}$, $p_{AI}=p_{RI}=0$ and so $p_A + p_P=1$. Inequality \ref{ineq:noLazy_2} yields $p_{I} \, L  + p_I \, L_e  \le p_{I} \, r_e + c_c $ which unfolds as $p_{I}\le \frac{c_c}{L+L_e-r_e}$, and inequality \ref{ineq:0_vs_e_3} yields $p_{I} \le (p_A+p_P) \frac{r_e}{L+L_e} = \frac{r_e}{L_e}$. The first condition is much stronger since $c_c \ll r_e \ll L_e$ so we only need $f \le \frac{c_c}{L+L_e-r_e} n$ for the equilibrium to hold.
    \end{itemize}
    Since in all those case the equilibrium holds only when Byzantines are too few to constitute a significant threat for the rational agents we consider that this is a leftover from the threat-less equilibria. In figure\ref{fig:Other_Equ}, we represented the equilibrium under the first two conditions with $\varepsilon_1=\frac{c_c}{L_e}$ and $\varepsilon_2=\frac{c_c}{L_e-c_c}$. The third condition being of measure $0$ in the simplex we did not represent it.
    
    \item Then for $\sigma_0$, valid blocks get $h$ endorsements. So from inequality \ref{ineq:participation_2}, we need $h<Q-1$ (which is equivalent to $f+g>n-Q+1$), otherwise $p_{AV}+p_{PV}>0$ and is non-negligible since honest agents propose valid blocks and $h \ge Q-1$. With this, we get $p_A+p_P\le p_I$ inequality \ref{ineq:0_vs_e_3} is satisfied since $p_{PI} \, L + p_I\, L_e  \ge p_I\, L_e$ and $(p_A+p_P) \, r_e \le p_I\ r_e$ and $L_e \gg r_e$.
    
    \item Finally for $\sigma_h$, valid blocks get $h+g$ endorsements, invalid ones get $f$. From inequality \ref{ineq:participation_2}, we get that we need $p_{AV}+p_{PV}>0$ so we need $h+g \ge Q$ (which is equivalent to $ f \le n-Q$). With this valid blocks get accepted, so rationals propose valid blocks. and this gives $p_{AV}+p_{PV}\ge \frac{Q}{n}>\frac{c_c}{r_e}$ (since we have $c_c \ll r_e$ and we do not have $Q \ll n$ ). This ensures that this inequality \ref{ineq:participation_2} is indeed verified. Then from inequality \ref{ineq:noLazy_2}, we would need $p_{PI} \, L  + p_I \, L_e  \ge (p_{AI} + p_{PI}) \, r_e + c_c$ for the equilibrium to hold, we distinguish two cases
    \begin{itemize}
        \item if $f < \frac{c_c}{L_e} n = \varepsilon_1$, since $Q$ is not small before $n$, since $L_e \gg c_c$ (so $f$ is small before $n$) and since invalid block receive $f$ endorsements, then invalid blocks are rejected. This means that $p_{AI}=p_{PI}=0$ and the inequality we needed becomes $p_I \, L_e  \ge c_c$ which is not satisfied (this is the hypothesis of this case) 
        \item Otherwise $f \ge \frac{c_c}{L_e} n = \varepsilon_1$. If $f$ is small before $n$, since $Q$ is not we have the same properties as before but this time the inequality is satisfied: the equilibrium holds. Otherwise $f$ is not small before $n$ so $p_I$ is not small, and since $L_e \gg r_e \gg c_c$ then $p_I L_e \ge r_e +c_c$. But since $p_{PI} \, L  + p_I \, L_e  \ge p_I \, L_e $ and  $r_e +c_c \ge (p_{AI} + p_{PI}) \, r_e + c_c$, it means that the inequality is satisfied.
    \end{itemize}
\end{itemize}


\subsection{With both the fines and trap blocks}

The idea of the trap blocks is to ensure that $p_I$ has an inferior bound $p_{prop}>0$. Let's look for the value $p_{prop}$ guaranteeing $u(\sigma_h) > u(\sigma_e)$ in inequality \ref{ineq:noLazy_2}, i.e., that the honest strategy yields a larger payoff than blind endorsement. Specifically, we need to guarantee that:

$$
p_{PI} \, L  + p_I \, L_e > (p_{AI} + p_{PI}) \, r_e + c_c
$$

Since: 

$$
p_{PI} \, L  + p_I \, L_e \ge p_I \, L_e \ge p_{prop} \, L_e
$$ 

and since:

$$
(p_{AI} + p_{PI}) \, r_e + c_c \le r_e + c_c
$$ 

Then to guarantee $u(\sigma_h) > u(\sigma_e)$ we can take:

\begin{equation}\label{ineq:noLazy_condition}
\begin{split}
p_{prop}>\frac{r_e+c_c}{L_e}
\end{split}
\end{equation}

Note that since $L_e$ is also a design parameter for the first amendment, there is a balance to find: if one want to reduce $p_{prop}$ then for this inequality to be respected, one could increase $L_e$. 
With this new property that ensures $p_I \ge p_{prop}>\frac{r_e+c_c}{L_e}$ we can lead the same reasoning as before to find the BARNE without looking for the ones for $\sigma_e$ as it is now strictly dominated by $\sigma_h$. 

\begin{itemize}
    \item For $\sigma_0$, the exact same reasoning as before holds, and we only need $f+g>n-Q+1$ for the equilibria to exist.
    \item For $\sigma_h$, we can follow the previous reasoning giving us $f \le n-Q$ with inequality \ref{ineq:participation_2}, but this time inequality \ref{ineq:noLazy_2} does not yield another constraint since we calibrated $p_prop$ for it to hold anywhere in the simplex.   
\end{itemize}

\section{Norms in the Byzantine-rational simplex}\label{appendix: Norm in simplex}

In Definition \ref{def:loc-stable}, the choice of the relevant norm $\Vert.\Vert_\nu$ is worthy of a small reflection. Amongst the usual norms, one could be tempted to choose the $\Vert.\Vert_\infty$, indeed this norm has the following advantage: the points $(f+x,g)$, $(f,g+x)$, $(f-x,g+x)$, are at the same distance from $(f,g)$ when $x\in \mathbbm{Z}$. These represent a single change in type of $x$ agents respectively from honest to byzantine, honest to rational, byzantine to rational. However, one might wish for the norm to be euclidean, but $\Vert.\Vert_2$ has the obvious problem of leading to a distance of $\sqrt{2}$ when $1$ rational becomes byzantine meaning we go from $(f,g)$ to $(f+1,g-1)$. 

In order to use a sensible euclidean norm, one needs to step back into the third dimension of the simplex where $(f,g)$ corresponds to the point $(f,g,h)=(f,g,n-f-g)$. So a more complex and Euclidean (but less intuitive) norm  could be: 

$$\Vert.\Vert_{2^*}:\ (f,g) \mapsto \frac{1}{\sqrt{2}}\Vert(f,g,n-f-g)\Vert_2$$

Where the $\sqrt{2}$ factor allows for the change of type of 1 agent to have a distance 1 from its point of origin no matter the change.

\end{document}